\documentclass[aps,prb,twocolumn,showpacs,10pt]{revtex4-1}
\usepackage{amsmath}
\usepackage{graphicx}
\usepackage{subfigure}
\usepackage{color}
\begin{document}
\title{Current-Driven Domain Wall Depinning and Propagation in Notched
Nanowires}

\author{H. Y. Yuan and X.R. Wang}
\email{[Corresponding author:]phxwan@ust.hk}
\affiliation{$^{1}$Physics Department, The Hong Kong University of
Science and Technology, Clear Water Bay, Kowloon, Hong Kong}
\affiliation{$^{2}$HKUST Shenzhen Research Institute, Shenzhen 518057, China}
\date{\today}

\begin{abstract}
Adiabatic spin transfer torque induced domain wall (DW) depinning
from a notch and DW propagation in a nanowire with a series of
notches is investigated. Surprisingly, notches help a current
to depin a DW and make a DW easier to propagate along a wire.
Following fascinating results on DW dynamics are found.
1) The depinning current density of a DW in a notch is 
substantially lower than the intrinsic threshold value below which 
a sustainable DW propagation doesn't exist in a homogeneous wire. 
2) The DW displacement from a notch is insensitive to notch 
geometry and current density when it is between the depinning 
and the intrinsic threshold current density. 
3) A current density below the intrinsic threshold value can 
induce a sustainable DW propagation along notched nanowires.
These findings not only reveal interesting and complicated
interaction between a current and a DW, but also have profound
implications in our current understanding of current-driven
DW dynamics as well as in the design of spintronic devices.
\end{abstract}

\maketitle

Controlled manipulation of domain walls (DWs) in magnetic
nanowires is an important topic in nanomagnetism not only
for its fundamental interest, but also for its applications in
spintronic devices \cite{Parkin1,Allwood,Chappert} where pinning,
depinning, DW displacement, and DW propagation are crucial. 
Magnetic fields via energy dissipation \cite{Walker,Energy} 
and electric current via angular momentum transfer 
\cite{Berger,SZhang1,Ono} are well-known DW control parameters. 
Although many efforts have been devoted to it and much progress 
has been made, our current understanding of the subject is still 
limited and far from satisfactory, especially for DW dynamics 
involving notches. The pinning field of a DW in a homogeneous 
nanowire is zero \cite{Walker,Energy}. To pin a DW in a wire, the 
wire inhomogeneity is necessary and notches are often used in 
experiments and simulations \cite{McMichael1,Klaui1,Parkin2,Bogart1} 
for positioning a DW. To move a DW out of a notch, a field 
larger than a critical value, called depinning field, is needed. 
The depinning field depends sensitively on notch properties 
as well as DW types \cite{Parkin2,Bogart1,yuan1}. 
It is so sensitive that depinning fields have been used to 
distinguished one type of DW from another in experiments 
\cite{Parkin2}.

Things are quite different when a current is used to manipulate DWs,
as shown by a mysterious observation \cite{Parkin2} that depinning 
current required to move a DW out of a notch does not depend on DW 
types, which is in a sharp contrast to its magnetic field counterpart. 
In principle, a current generates an adiabatic spin transfer torque 
(STT) \cite{Berger} and a non-adiabatic STT \cite{SZhang1,Thiaville}.
The mechanism of the adiabatic torque is well established while
the non-adiabatic torque is still in debate \cite{Seo, Koyama}.
The non-adiabatic STT, even existing, is much smaller than the adiabatic 
one, and it is neglected in many theoretical treatments and analysis. 
In the absence of the non-adiabatic STT, a current density below a 
threshold value can only displace a DW for a finite distance when the 
current is on and the displacement vanishes when the current is off 
\cite{SZhang2,Tatara}.
The transit displacement, which depends on material properties,
is on the order of a few hundreds nanometers for permalloy.
Only under a current density above the threshold value, the DW 
can propagate continuously\cite{SZhang1,Thiaville,Parkin3}.
This phenomenon is called the intrinsic pinning for the adiabatic STT.
How an adiabatic STT interacts with a DW trapped in a notch or how the
adiabatic STT drives a DW to propagate along a wire with many notches
is obviously important \cite{Parkin1,Allwood}, but little known.
Naively, one may expect that a DW would be further pinned by a notch.
Thus, the nature conjecture is that a current density above the threshold 
value for the homogeneous wire would be required to move the DW 
out of the notch. It would also be nature to expect that a larger 
current density is required to induce a sustainable DW propagation 
along the wire with many notches. In this letter, we numerically 
study the issue. Surprisingly and fascinatingly, a current density 
substantially below the threshold value can move a DW out of a notch.
The DW is far from the notch when the current is switched off.
In fact, the displacement, inversely proportional to the damping and
proportional to DW width, is much larger than both the DW width and the
transit displacement of the DW in the corresponding homogeneous wire. 
If one places a series of notches along the wire, a current density 
much smaller than the intrinsic threshold value can also sustain 
a continuous DW propagation.

\begin{figure}
\centering
\includegraphics[width=0.4\textwidth]{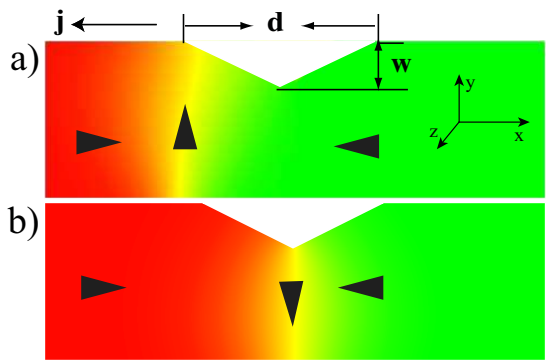}\\
\caption{(color online) Side view of a notched nanowire with an
anticlockwise transverse wall (a) and a clockwise transverse wall (b).
The color indicates the magnitude of $m_x$, varying from green for $m_x
= -1$ to red for $m_x = +1$ with yellow for $m_x=0$. The thick arrows
indicate the magnetization directions. $\mathbf{j}$ is in the -$x$-direction. }
\label{fig1}
\end{figure}

We consider magnetic wires of 5000 nm long, 4 nm thick, and width $W$ 
varying from 48 nm to 120 nm so that transverse DWs are preferred.
A triangular/rectangular notch of width $d$ and depth $w$ is 
located on the top edge of the wire as shown in Fig. \ref{fig1}.
The $x-$, $y-$, and $z-$axis are respectively along the length, 
width, and thickness directions with the origin at the wire center. 
The magnetization dynamics is governed by the generalized 
Landau-Lifschitz-Gilbert (LLG) equation,
\begin{equation}
\frac{\partial\mathbf{m}}{\partial t}=-\gamma\mathbf{m\times H}_{
\mathrm{eff}}+\alpha\mathbf{m}\times\frac{\partial\mathbf{m}}{\partial t}
-(\mathbf{u} \cdot \mathbf{\nabla} )\mathbf{m},
\end{equation}
where $\mathbf{m}$, $\gamma$, $\alpha$, and $\mathbf{H}_{\mathrm{eff}}$
denote the unit vector of magnetization, gyromagnetic ratio, the Gilbert
damping constant and effective field, respectively.
$\mathbf{H}_{\mathrm{eff}}$ includes exchange field and anisotropy field.
The third term on the right hand side is the adiabatic STT, where
$u=jP\mu_B/(eM_s)$ is proportional to the current density $j$ and is
along the current direction, where $P$, $\mu_B$, $e$ and $M_s$ are the 
current polarization, the Bohr magneton, the electron charge and the 
saturation magnetization, respectively. In our simulations, electrons 
flow in the $+x$-direction (opposite to $\mathbf{j}$ illustrated in 
Fig. \ref{fig1}). Thus $\nabla$ in this term becomes $\partial/\partial x$.
The LLG equation is solved numerically by {\footnotesize OOMMF}
package \cite{oommf}. To mimic permalloy wires, we use the exchange
constant $A = 1.3 \times 10^{-11}$ J/m, $M_s = 8 \times 10^5$ A/m,
zero crystalline anisotropy and the damping constant $\alpha = 0.02$.
The mesh size is 4nm$\times$4nm$\times $4nm.

\begin{figure}
\centering
\includegraphics[width=0.45\textwidth]{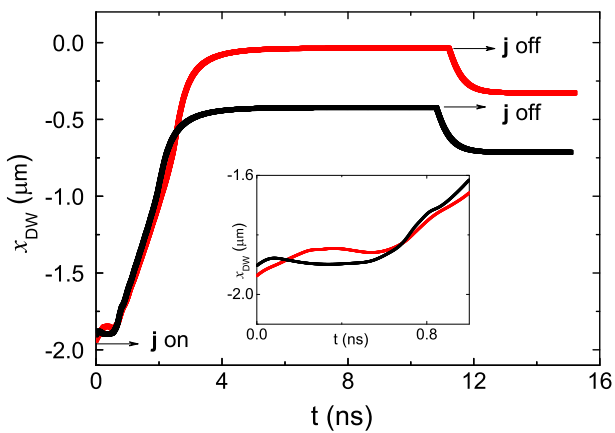}\\
\caption{(color online) The time evolution of $x_{DW}$, defined as 
the average x-coordinate of those spins with $m_x=0$.
The wire width is 64 nm, and the notch of $w=16$ nm and $d=64$ nm is 
located at -1900 nm. The red (black) curve is for an AW (CW) under 
current density of $u = 650$ m/s. The inset is the enlarged part of 
$x_{DW}(t)$ for $t<1$ ns. }
\label{fig2}
\end{figure}

All wires show similar behaviors, and we present the results for a 
64 nm wide wire. There are two types of transverse DWs, anticlockwise 
wall (AW) which prefers to stay near the notch edge as
sketched in Fig. \ref{fig1}a and clockwise wall (CW) which tends to
reside near the notch center as sketched in Fig. \ref{fig1}b.
The intrinsic threshold current density of the corresponding uniform 
nanowire is $u_c = 700$ m/s only above which a DW can undergo a 
sustainable propagation. In the presence of a triangular notch of 
$w=16$ nm and $d=64$ nm centered at $x=-1900$ nm, a current density 
of 578 m/s $\leq u \leq u_c$ can obviously depin an AW from the notch. 
The time evolution of the AW position under $u = 650$ m/s, 
on at $t=0$, is plotted as the red curve in Fig. \ref{fig2}.
The AW, pinned at the left edge of the notch initially, moves first 
in the $+x-$direction and reaches the right side of the notch in about 
0.2 ns. Its center $x_{DW}$, defined as the average $x$-coordinate
of those spins with $m_x=0$, does not change much for about 0.4 ns, 
shown in the enlarged figure in the inset while the DW structure
continuously deforms and an antivortex is born at the edge defect 
of winding number \cite{yuan2} -1/2 on the lower wire edge.
Right after the birth of the antivortex, the DW starts to move out of 
the notch at a constant velocity represented by the linear segment 
between 0.6 ns and 3.5 ns. The DW stops finally at $x_{DW}=-36$ nm, 
almost 2 $\mu$m away from the notch. Intuitively, one will expect 
that notches tend to strengthen DW pinning so that any current 
density (measured by $u$) below $u_c$ would not depin a DW. 
Thus, what we observed is a very surprising result.
When the current is switched off at 11.2 ns, as expected \cite{SZhang2},
the DW retreats for about 292 nm, the intrinsic transit displacement.
At the end of the process, the DW displacement is about 1572 nm which 
is far away from the notch. This displacement is more than five times
of the transit displacement of a DW under the same current density if
the wire was uniform. Of course, different from our results, the net
DW displacement would be zero n a uniform wire after the current is 
switched off \cite{SZhang2}.

The overall CW behavior is very similar to the AW case as shown
by the black curve in Fig. \ref{fig2} for $u = 650$ m/s ($<u_c$).
The DW center does not change much initially (black line in the inset),
probably because the center is already in the right side of the notch,
while the DW structure deforms for about 0.6 ns before a vortex appears 
at the edge defect of winding number 1/2 on the lower wire edge.
Similar to the AW case, the DW center starts to move out of the notch,
at a constant velocity, after the vortex is born. It stops at $x_{DW}
= -422$ nm far out of the notch. When the current is switched off at
$t=10.8$ ns, the DW center retreats for about $292$ nm and stops
at $x_{DW} = -714$ nm, about 1 $\mu$m on the right of the notch.
There are some subtle differences in comparison with the AW case. 
An antivortex may also generated at the edge defect of winding number
$-1/2$ \cite{yuan2}. The final DW displacement depends also on whether
an antivortex or a vortex is generated. The details of how many 
depinning process there are, how a DW is depinned in each process, and 
what is the phase diagram in the $u-d$ plane will be reported elsewhere.

\begin{figure}
\centering
\includegraphics[width=0.45\textwidth]{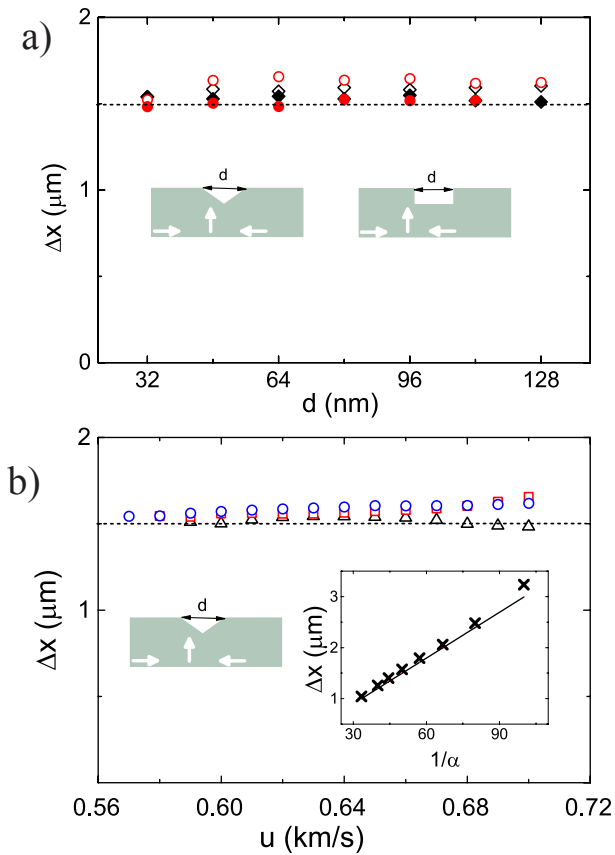}\\
\caption{(color online) $\Delta x $ for an AW in a wire of 64 nm wide  
with a triangular (open symbols) or a rectangular (filled symbols) notch.
a) $\Delta x $ vs. $d$. Current density is $u = 650$ m/s (diamonds) 
and 700 m/s (circles). The insets illustrate the notches used. 
b) $\Delta x $ vs. $u$ with notch width $d =$32 nm (triangles), 64 nm 
(squares), and 128 nm (circles). The dotted lines are Eq. (3) without 
any fitting parameters. The left inset: Notch used. The right insect: 
$\Delta x $ vs. $\alpha$ (crosses for simulations and the line 
for Eq. (\ref{dx})). }
\label{fig3}
\end{figure}

It is interesting to know how the DW displacement depends on $u$ (larger than
depinning value $u_d$ and smaller than $u_c$), the notch type and geometry.
Let $\Delta x$ be the net DW displacement, i.e. the displacement after 
the current is switched off. Fig. 3a is $d$-dependence of $\Delta x$ of
an AW with rectangular (filled symbols) and triangular (open symbols) notches
for $u= 650$ m/s (diamonds) and $700$ m/s (circles). Surprisingly, $\Delta x$
does not depend on the notch types and geometry, within numerical accuracy.
Fig. 3b is $u$-dependence of $\Delta x$ for triangular notches of various
widths $d=32$ nm (triangles), 64 nm (squares), and 128 nm (circles).
$\Delta x$ depends only weakly on $u$.
It is interesting to notice that depinning occurs only when a
vortex/antivortex is generated at one of the DW edge defects. 
DW moves, together with the vortex/antivortex. The DW displacement 
is the distance that the vortex/antivortex travels in the $x-$direction 
in its lifetime. If this observation is the essential depinning physics, 
the highly counter-intuitive DW displacement behavior can be understood 
from the Thiele equation \cite{Thiele,Thiaville,Huber} for a vortex.
Using the Thiele equation \cite{Thiele,Thiaville,Huber}, one has
\begin{equation}
\begin{aligned}
\mathbf{F}+\mathbf{G}\times(\mathbf{v-u})+\mathbf{D}\cdot(\alpha\mathbf{v})=0
\end{aligned}
\end{equation}
where $\mathbf{F}$ is the force acted on the vortex core from external field
that is zero in our case, $\mathbf{G}$ is gyrovector, $-M_s/\gamma 2 \pi
q p l \mathbf{\hat{z}}$, where $q$ is the winding number (+1 for a vortex
and -1 for an antivortex) and $p$ is the vortex polarity ($\pm 1$ for core
polarization in $\pm z$ direction) and $l$ is the thickness of nanowire.
$\mathbf{v} = (\dot{x}_c, \dot{y}_c)$ is the velocity of the vortex core,
where $(x_c,y_c)$ is the core position.
$\mathbf{D}$ is dissipation dyadic, whose none zero elements are $D_{xx} =
D_{yy}=-2M_sWl/(\gamma \Delta)$ \cite{Huber}, where $\Delta$ is the
Thiele's DW width \cite{Thiele}. Then the $x-$component of the Thiele 
equation is $$-G_z \dot{y}_c + \alpha D_{xx} \dot{x}_c = 0.$$
The displacement of the vortex core can be obtained by integrating
the above equation, and one has
\begin{equation}
\Delta x = \pi \Delta/ \alpha .
\label{dx}
\end{equation}
Note that vortex travel distance in $y-$direction is $W$. 
Thus, we show that $\Delta x$ does not depend on $u$ and notch geometry! 
Eq. (3) (dashed lines without any fitting parameters) explains 
the numerical results very well as shown in Fig. 3. 
The $\alpha$ dependence is also confirmed numerically as shown in the
right inset of Fig. 3b, where crosses denote simulation results and
line is Eq. (\ref{dx}). Although not shown explicitly, the displacement 
of a CW, whose Thiele DW width is smaller than that of a AW, can also 
be explained by Eq. (3) equally well. 

\begin{figure}
  \centering
  \includegraphics[width=0.45\textwidth]{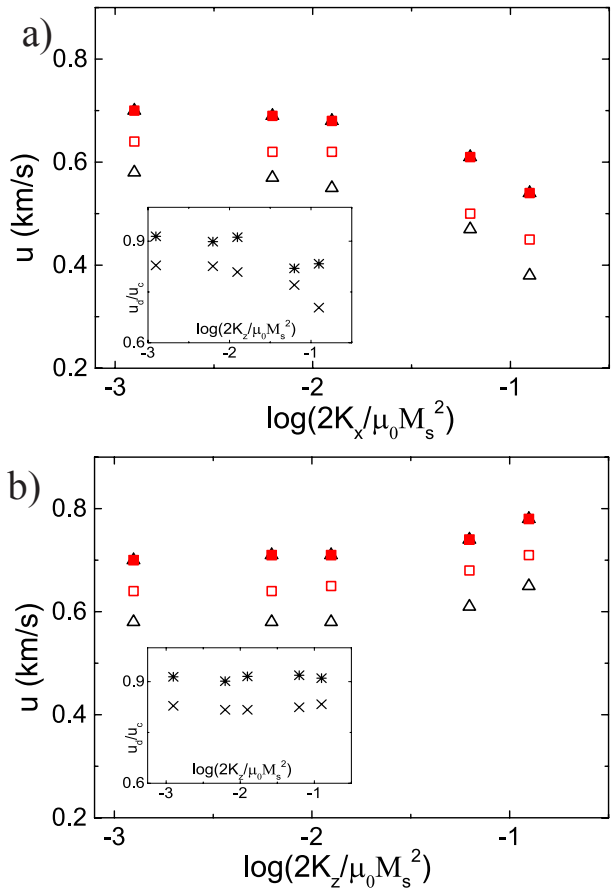}\\
\caption{The easy-axis anisotropy dependence a) and hard-axis
anisotropy dependence of the intrinsic threshold current density
$u_c$ (filled symbols) and the depinning current density $u_d$
(open symbols) for AW (triangles) and CW (squares).
The insets are $u_d/u_c$ vs. $K_x$ (a) and $K_z$ (b) 
for CW (stars) and AW (crosses).}\label{fig4}
\end{figure}

To demonstrate the robustness and generality of the results, we add an  
extra biaxial crystalline anisotropy of $(K_z m_z^2-K_x m_x^2)/2$ in 
our simulations, where $K_x$ and $K_z$ are respectively the easy- and 
hard-axis anisotropy coefficients. Fig. 4 are the $K_x$ ($K_z$) 
dependence of $u_c$ (filled symbols) and depinning current density $u_d$ 
(open symbols) for AW (triangles) and CW (squares) when $K_z=0.5\times 
10^3$ $\mathrm{J/m^3}$ ($K_x=0.5 \times 10^3$ $\mathrm{J/m^3}$). 
All other parameters are the same as those for Fig. 2. 
Fig. \ref{fig4} shows that the larger the easy-axis (hard-axis) 
anisotropy is, the smaller (larger) $u_c$ will be, which is 
consistent with the prediction of a biaxial model \cite{SZhang2}.
The depinning current density $u_d$ is well below the intrinsic
threshold current density $u_c$ as $K_x$ ($K_z$) changes. 
It is interesting to see that an extra easy-axis anisotropy reduces 
$u_d/u_c$ while $K_z$ hardly affects $u_d/u_c$, as illustrated 
in the insets of Figs. \ref{fig4}a and \ref{fig4}b, respectively.

The above results imply that notches should help current-driven DW
propagation along a wire instead of hindering the propagation as
in the case of field-driven DW motion. To confirm this highly
counter-intuitive conjecture, we place a series of triangular
notches (of 64 nm wide and 16 nm deep) along the two wire edges
alternatively, as illustrated in the low-right inset of Fig. 5.
The reason that notches are alternatively placed on the upper and lower
edges is because an AW (CW) will change to a CW (AW) after depinning
from a notch under the assistance of generated vortices/antivortices
on one wire edge that travel to the other edge and die there.
Our micromagnetic simulations show that a current density below $u_c$ 
can induce a sustainable DW propagation when notch separation is chosen 
properly according to DW displacement from a single notch discussed early. 
Fig. 5 is the $u-$dependence of average DW speed $\bar{v}$ (crosses) 
when the notch interval is 1800 nm for a wire of 64 nm wide. 
The average speed is zero for $u<578$ m/s because it is smaller than 
the depinning value. When the current density increases to $u \geq 578$ 
m/s, but well below $u_c = 700$ m/s, the DW has a speed of about 500 m/s.
The time-dependence of the instantaneous speed at $u=650$ m/s is given
in the top-left inset. The DW speed is almost zero at notches (red
bars) and is bigger than 500 m/s away from the notches.
As the current density increases to $u>u_c$, the average speed approaches
the DW speed in the corresponding homogeneous wire (red squares).
Without a surprise, notches cause the instantaneous DW speed
oscillating as shown in the low-left inset at $u=800$ m/s. 
Interestingly, the instantaneous DW speed at notches (red bars) 
is near the maximal in contrast to near zero speed for $u<u_c$.
\begin{figure}
  \centering
  \includegraphics[width=0.45\textwidth]{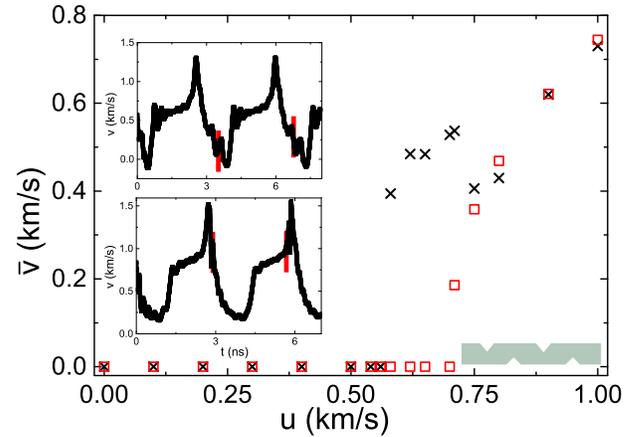}\\
\caption{$\bar{v}$ vs. $u$ for a notched nanowire (crosses) 
and a homogeneous nanowire (red squares) of 64 nm wide. 
The low-right: Notched nanowire used.
The top-left and low-left insets: The instantaneous DW 
speed for $u = 650$ m/s and $u = 800$ m/s, respectively.
The red bars indicate the moments when a DW is at notches.}\label{fig5}
\end{figure}

The findings presented here should have profound implications on 
STT-based DW applications as well as our understanding of STT-driven 
DW dynamics. For example, the original argument for the inclusion 
of the non-adiabatic STT was based on the experimental observation 
of DW propagation below the intrinsic threshold current density. 
The argument relies on an implicit assumption that any wire 
inhomogeneity shall always increase the depinning current density.
Thus, the experimentally observed DW propagation below the intrinsic
threshold current density could only be explained by including the 
non-adiabatic torque. Our findings obviously shake that reasoning.  
One needs to reexamine the analysis of necessity of the non-adiabatic 
STT, especially its magnitude.

The contrasting differences between current and field driven DW 
depinning and DW propagation in notched wires come from different 
control mechanisms. On the one hand, a magnetic field creates energy 
density difference between two domains that are separated by a DW.
According to Ref. 5, a static DW cannot exist between such two domains 
if the wire has the translational symmetry. This is why a DW cannot 
resist to an arbitrary small field. To prevent a DW motion under a 
field, the translational symmetry has to be destroyed. This could be 
done by inevitable wire roughness or by intentionally designed notches. 
On the other hand, a current exerts a torque on a DW through angular 
momentum transfer. As a result, a DW can deform its structure to 
absorb the current-generated STT, resulting in the intrinsic 
threshold current density in a homogeneous wire \cite{SZhang2}.
The introduction of a notch may weaken the delicate balance between 
DW structure deformation and adiabatic STT. This may be the origin 
of our counter-intuitive results. It should be interesting to 
explore this idea.

In conclusion, we have investigated the adiabatic STT-driven DW
depinning and DW propagation in a notched nanowire. Below the intrinsic
threshold current density, a DW can be depinned from a notch and be
displaced by a long distance that is inversely proportional to the
damping constant. In a realistic material, the displacement can be of
the order of $\mu$m. Furthermore, the DW displacement does not depend
on the notch type and geometry. It depends also very weakly on the
current density as long as the current density is between the 
depinning and the intrinsic threshold current density. This surprisin
result can be explained well by the Thiele equation for vortex dynamics. 
These findings should have profound implications in both STT-driven 
DW motion and STT-based DW devices.

This work was supported by the NSF of China
grant (11374249) and Hong Kong RGC grant (605413).
HYY acknowledges the support of Hong Kong PhD Fellowship.

\end{document}